# Ferrohydrodynamic Microfluidics for Bioparticle Separation and Single-Cell Phenotyping: Principles, Applications, and Emerging Directions

Yuhao Zhang[a], Yong Teng[g], Kenan Song[d], Xianqiao Wang[d], Xianyan Chen[e], Yuhua Liu[h],

Yiping Zhao[f], He Li[a], Leidong Mao*[c], Yang Liu*[a,b]

[a]School of Chemical, Materials and Biomedical Engineering, College of Engineering, The University of Georgia, Athens, Georgia 30602, USA

[b]Institute of Bioinformatics, The University of Georgia, Athens, Georgia 30602, USA

[c]School of Electrical and Computer Engineering, College of Engineering, The University of Georgia, Athens, Georgia 30602, USA

[d]School of Environmental, Civil, Agricultural, and Mechanical Engineering, The University of Georgia, Athens, Georgia 30602, USA

[e]College of Public Health, The University of Georgia, Athens, Georgia 30602, USA

[f]Department of Physics and Astronomy, The University of Georgia, Athens, Georgia 30602, USA

[g]Department of Hematology and Medical Oncology, Winship Cancer Institute, Emory University, Atlanta, Georgia 30322, USA

[h]Center for Disease Control and Prevention of Jinan Railway Bureau, Jinan, Shandong 2500002, China

*Email: Yang Liu (liuy@uga.edu); Leidong Mao (mao@uga.edu)

**Abstract**

Ferrohydrodynamic microfluidics relies on magnetic field gradients to manipulate diamagnetic particles in ferrofluids-filled microenvironments. It has emerged as a promising tool for label-free manipulation of bioparticles, including their separation and phenotyping. This perspective reviews recent progress in the development and applications of ferrofluids-based microfluidic platforms for multiscale bioparticle separation, ranging from micron-scale cells to submicron extracellular vesicles. We highlight the fundamental physical principles for ferrohydrodynamic manipulation, including the dominant magnetic buoyancy force resulting from the interaction of ferrofluids and particles. We then describe how these principles enable high-resolution size-based bioparticle separation, subcellular bioparticle enrichment, and phenotypic screening based on physical traits. We also discuss key challenges in ferrohydrodynamic microfluidics from the aspects of ferrofluids' biocompatibility, system throughput, and nanoparticle depletion. Additionally, we outline future research directions based on the integration of machine learning, 3D printing, and multiplexed detection. Together, these insights outline a roadmap for advancing ferrofluids-based technologies in precision biomedicine, diagnostics, and cellular engineering.

**Introduction**

Single-cell analysis is rapidly gaining attention due to its transformative potential in resolving cellular and molecular heterogeneity and in uncovering mechanisms underlying human biology and disease[1-3]. Accurate sorting of cells and their phenotypic profiling are critical for downstream molecular analyses, enabling researchers to link cell states with transcriptional, proteomic, or functional profiles[4-10]. However, these analyses face significant challenges from background noise introduced by contaminating or non-target cells. The purity and quality of input cell samples directly impact processing costs, data complexity, and the reliability of downstream single-cell measurements, particularly when studying rare populations, such as circulating tumor cells (CTCs) in patient blood [3, 5, 11, 12].

Achieving high-throughput, high-purity enrichment of target cells is essential not only for single-cell research but also for broader biological applications such as bacterial detection in food safety, tumor-derived exosome isolation, and stage-specific T cell preparation for CAR-T cell manufacturing[13-15]. However, existing biomedical sorting systems often fail to meet these growing demands. Current approaches are limited by high cost, low throughput, insufficient resolution, and the inability to manipulate nanoscale targets or perform multimodal sorting. Traditional separation methods, such as magnetic-activated cell sorting (MACS) and fluorescence-activated cell sorting (FACS), rely heavily on molecular labeling and expensive reagents[16, 17]. In contrast, microfluidic technologies have emerged as powerful tools for bioparticle separation and purification due to their precise sample handling and integration capabilities. For example, a variety of microfluidic platforms, including acoustophoresis[18], dielectrophoresis[19-21], dean flow[22-24], and ferrohydrodynamics[15, 25, 26], have been explored for label-free separation of CTCs and other bioparticles.

Among these, ferrohydrodynamic manipulation has recently gained increasing attention. Ferrohydrodynamics refers to the study and application of fluid behavior in ferrofluids, colloidal suspensions of magnetic nanoparticles, under the influence of an external magnetic field[27]. When exposed to external magnetic fields, ferrofluids exhibit tunable magnetic and flow properties, enabling new modes of particle manipulation within microfluidic environments[28]. While originally developed for macroscale engineering applications such as sealing[29], damping[30], mixing[31], and heat transfer[32], ferrofluids are now being adapted to microscale biology due to their unique ability to facilitate label-free, high-resolution control of bioparticles. Ferrohydrodynamic platforms represent a convergence of disciplines, combining physics, microfluidics, nanomaterials, and biology, and offer compelling advantages in continuous particle sorting and high-resolution screening based on physical or mechanical traits. Over the past decade, researchers, including our lab, have demonstrated that ferrohydrodynamic microfluidic systems can manipulate particles across a wide size range, from submicron extracellular vesicles to micron-scale mammalian cells, without the need for molecular labels.[26, 33-36] These advances position ferrohydrodynamics as a highly promising solution for emerging biomedical applications, including rare cell enrichment, single-cell phenotyping, and liquid biopsy development—domains where conventional methods often suffer from limited sensitivity, low throughput, and high operational costs.

In this perspective, we review recent advances in ferrofluids-based ferrohydrodynamic manipulation for biological particle separation and cell phenotyping (Figure 1). We begin by reviewing the underlying physical principles and then explore current implementations across size scales and cell traits. We also discuss key challenges and future directions, aiming to provide a roadmap for advancing ferrohydrodynamics-enabled microfluidic technologies in precision biomedicine and bioengineering.

**Principles of Ferrohydrodynamic Manipulation**

Ferrohydrodynamic particle manipulation utilizes unique negative magnetophoresis in ferrofluids-filled microfluidic systems to achieve label-free control of particles and cells. Unlike positive magnetophoresis[37, 38], which requires magnetic labeling of cells with antibody-bead complex, negative magnetophoresis exploits the intrinsic magnetization contrast between diamagnetic particles (e.g., CTCs) and their surrounding paramagnetic or superparamagnetic medium[39, 40]. In such systems, biological particles behave as magnetic voids and experience a repulsive force when exposed to a non-uniform magnetic field[41]. The primary driving force in ferrohydrodynamic manipulation is the magnetic buoyancy force, which displaces diamagnetic particles toward regions of weaker magnetic fields.

$$\mathbf{F_{mag}} = \mu_0 V_p (\mathbf{M_p} - \mathbf{M_f}) \cdot \nabla]H \qquad (1)$$

Here, $\mu_0$ is the magnetic permeability of free space. $V_p$ is the volume of the diamagnetic particle, $\mathbf{M_p}$ and $\mathbf{M_f}$ are the magnetization of the particle and ferrofluids, respectively, and $\mathbf{H}$ is the applied magnetic field strength at the center of the particle. When a magnetic field gradient is applied, the ferrofluids nanoparticles are attracted toward the region with field maxima, while the diamagnetic particles are effectively displaced toward the field minima, acting as "magnetic holes" in the fluid[27]. Since the magnitude of this force scales with the cube of the particle diameter, ferrohydrodynamic systems are particularly well-suited for high-resolution size-based particle manipulation. Typical ferrohydrodynamic systems consist of microfluidic channels paired with external magnetic sources, such as permanent magnets, pole arrays, or electromagnets, to generate non-uniform magnetic fields. The design and positioning of the magnets determine the magnetic flux density and field gradient, which, along with the ferrofluids' composition, govern manipulation efficiency.

In addition to the magnetic buoyancy force ($F_{mag}$), several other physical forces influence the particle behaviors in these microfluidic ferrohydrodynamic systems, including gravitational ($F_{gravity}$) and buoyant forces ($F_{buoyancy}$) and hydrodynamic viscous drag force ($F_{drag}$), along with Brownian motion, which becomes increasingly significant in submicron regimes. The net motion of a particle suspended in the ferrofluids can be described by the balance of these forces (Equation 2):

$$F_{net} = F_{mag} + F_{gravity} + F_{buoyancy} + F_{drag} \quad (2)$$

The manipulation resolution and throughput are determined by key parameters, including particle size and density, ferrofluids viscosity, magnetic flux density and gradient, and microchannel geometry. By specifically tuning these variables, ferrohydrodynamic systems can be adapted for a range of bioparticle manipulations, including high-throughput rare cell separation, subcellular particle sorting, and biophysical single-cell phenotyping. In the following sections, we review how these principles have been applied to design ferrofluids-based microfluidic platforms for a range of biological targets and scales: from micron-sized cells to submicron extracellular vesicles and physical phenotype-based cell analysis.

**Applications in Micron-scale Particles Separation**

Micron-scale bioparticles, including cells, platelets, and bacteria, are essential for both fundamental and biological research, clinical diagnostics, and food safety, and these particles often carry critical disease-relevant information. For example, CTCs are key mediators of cancer metastasis[42]; T lymphocyte subpopulations influence immunotherapy outcomes[43]; Platelets are involved in thrombosis and cancer progression[44]; and bacteria are major contributors to infectious diseases and microbiome-related health conditions[45].

Despite their importance, the efficient manipulation and separation of micron-scale particles, particularly rare targets, from large volumes of biological samples remains a significant challenge. Achieving a high-throughput and high-resolution separation is essential for the processing of large-scale samples. Furthermore, enriching target particles with high purity not only enhances data reliability but also reduces downstream processing costs, yet doing so in a scalable and reproducible manner continues to be difficult. The choice of cell separation technique, whether label-free or label-based, is often dictated by the properties of the target particles. However, challenges such as low abundance, morphological heterogeneity, variability in surface marker expression, and particle fragility make enrichment difficult, especially for rare cell populations like CTCs[46].

Ferrohydrodynamic separation, when integrated into microfluidic platforms, offers a promising solution for label-free, size-based enrichment. This approach exploits the size-dependent magnetophoretic behavior of bioparticles suspended in ferrofluids, providing a cost-effective and equipment-minimal method for separation. For instance, Ayse et al. applied ferrofluids in *E. Coli* separation with a trapping efficiency of 93.7% and purity of 76.1%[33]. Xuan et al. demonstrated the use of ferrofluids to separate yeast cells based on size[34]. Our lab has employed ferrofluids to perform size-based separation for various cell types, including HeLa cells[47], human cancer cells[36], CTCs[15, 48], and T cells[15]. To enhance resolution and throughput, we designed an inertial-enhanced ferrohydrodynamic platform, inertial-FCS, which combines inertial focusing with ferrohydrodynamic separation[15]. This system achieves a size resolution as fine as 1 µm in diameter and has been successfully applied to isolate small lymphocytes (7-8 µm) from peripheral blood mononuclear cells (PBMCs). Lin et al. employ this size-based method to deplete microscale particles from the blood to generate plasma[35]. Furthermore, Xuan et al. investigated the feasibility of shape-based separation using diamagnetic particles, paving the way to further cellular morphology-based separation via ferrofluids[49].

Nonetheless, size overlap between target and background cells can lead to sample loss, particularly when isolating rare populations. CTCs, for example, are present at only 1-10 cells per mL of blood and, although typically larger than white blood cells (WBCs), can exhibit substantial size overlap[50, 51]. To address these limitations, we developed a size-inclusive, antigen-independent platform - iFCS[52]. This method integrates magnetophoresis to remove magnetically labeled WBCs using antibodies and magnetic beads, while utilizing diamagnetophoresis to enrich unlabeled target cells. The iFCS system has achieved near - 100% recovery rates with enhanced purity. Additionally, by incorporating inertial focusing, the iFCS strategy is capable of processing over 60 mL of whole blood per hour, corresponding to approximately $2 \times 10^5$ cells per mL[15].

However, concerns regarding the biocompatibility of ferrofluids remain[53-55]. Cells can internalize ferrofluids nanoparticles, potentially disrupting essential biological functions such as signaling, gene expression, and metabolism. Most commercial and custom-synthesized ferrofluids are not suitable for long-term storage or incubation of live cells, as prolonged exposure can compromise cell viability and function. In our lab, we have developed a biocompatible ferrofluids formulation that minimizes adverse effects on cell viability, short-term proliferation, and functional behaviors such as motility[36]. Nevertheless, minimizing the interaction time between cells and ferrofluids remains essential. Thus, achieving reduced processing time and maximizing throughput are critical design criteria for ferrohydrodynamic separation systems, particularly when handling large volumes of primary or clinical samples.

**Submicron Particle Separation**

Submicron particles ranging from 10 nm to 1000 nm, including exosomes, microvesicles, and apoptotic bodies, are attracting increasing attention due to their importance in intercellular communications, disease diagnostics, therapeutic delivery, and understanding of pathological processes[56-58]. Conventional methods (e.g., ultracentrifugation) for separating these particles often suffer from limitations such as low throughput, high sample loss, long processing times, and difficulty in isolating particles with significant size overlap[59]. While microfluidic techniques offer promising capabilities for precisely manipulating submicron particles using external forces (e.g., acoustic waves), there are inherent challenges[60-62]. As particle size decreases, Brownian motion becomes increasingly significant while the magnitude of most externally applied forces scales down. This mismatch reduces control over particle trajectories, making it difficult to achieve high-resolution, high-throughput separation at the nanoscale.

To address these challenges, our lab first applied negative electrophoresis on exosome separation using a new platform – FerroChip[26]. We utilized a quadrupole magnet array capable of generating a magnetic flux density of up to 0.52 T and a magnetic flux density gradient of up to 1272 T/m. These high magnetic field conditions make it possible to manipulate nanoscale particles using the magnetic buoyancy force. FerroChip demonstrated efficient focusing and enrichment of exosome-like nanoparticles, achieving a recovery rate of 94.3%, purity of 87.9%, and a throughput of 60 μL/h. Lin et al. engineered an ultra-high-gradient magnetic field module by integrating permanent magnets, high permeability perm-alloys, and on-chip magnetic pole arrays[35]. This system can generate a magnetic flux density gradient of up to 100,000 T/m, improving the separation performance. When applied to the separation of small vesicles (sEVs), the platform achieved a recovery rate of 85.8% and purity of 80.5%. While these early studies highlight the potential of ferrohydrodynamic separation in submicron regimes, broader applications (i.e., virus isolation) remain under development. Several challenges must still be addressed, including the strict requirement for the super-strong magnetic field, complex magnet setup, limited throughput, and the difficulty of depleting residual ferrofluids nanoparticles post-separation. Overcoming these obstacles is essential for advancing ferrofluids-enabled submicron particle separation into more widespread biomedical and clinical applications.

**Cell Phenotyping Using Ferrofluids**

Cell phenotyping based on physical and mechanical properties, such as size, shape, density, deformability, and surface antigen-binding capacity, offers insights into cellular heterogeneity, state, and function. Investigation of these traits at the single-cell level has significant value in the study of a wide range of biological processes, including cancer metastasis, immune response, and disease progression[63-67]. For example, tumor cells undergoing epithelial-to-mesenchymal transition (EMT) display altered size and deformability, contributing to enhanced invasiveness[68]; immune cell subpopulations can be categorized based on the surface antigen expression[69]; and cell density has been shown to correlate with metabolic activity, as observed in tumor cells under stress of hypoxia[70]. Despite their importance, single-cell or near-single-cell level phenotyping of these traits remains challenging. Methods such as flow cytometry, atomic force microscopy (AFM), and density-gradient centrifugation often lack the sensitivity to detect subtle differences in physical traits, suffer from low throughput, or are not well-optimized for single-cell resolution[71-73].

Ferrofluids-based microfluidics offers a scalable, sensitive, and label-free platform to interrogate these phenotypes with a high resolution. By balancing magnetic buoyancy force and other forces (e.g., gravitational force, buoyancy force, hydrodynamic drag force, magnetic force) in the microfluidic channel, ferrofluids enable continuous, real-time manipulation and analysis of cell populations based on cell phenotypes. Durmus et al. introduced a magnetic levitation platform, MagDense, capable of measuring single-cell density with a precision down to $1 \times 10^{-4}$ g/mL using a paramagnetic medium[74]. This platform has been applied in various cancer subtypes and has monitored the dynamic density changes in response to environmental stress and antibiotic treatment. Knowlton et al. employed the magnetic levitation principle to observe sickle cells in blood[75]. The underlying principle in these systems is based on force equilibrium: the net vertical force on each cell is determined by the balance of gravitational, buoyant, and magnetic buoyancy forces. The resulting equilibrium height in the fluid column is directly linked to the cell's density. Our lab has recently designed two complementary ferrohydrodynamic systems for physical phenotyping. Inertial-FCS can precisely manipulate cells based on the size differences with a precision of 1 μm in diameter[15]. The system can resolve and sort heterogeneous cell populations into five discrete size-defined subpopulations, each with a near single-cell size distribution. In parallel, qFCS expands the modality to surface marker profiling[76]. By conjugating magnetic beads to cell surface antigens, qFCS can sort cells based on the surface antigen-binding capacity and estimate the surface marker density with high accuracy. This phenotyping method is governed by the magnetization contrast between the ferrofluids and labeled cells, which is directly correlated to the number of beads bound and, thus, to the level of surface antigen expression. The platform supports rare cell enrichment (down to 10 cells/mL) and low-abundance antigen quantification, as demonstrated in profiling CD154 expressing on activated and non-activated T cells. Together, these ferrofluids-based platforms provide a multidimensional view of cell phenotypes at single-cell resolution. By enabling continuous, high-throughput, and label-free investigation of physical traits, they offer powerful tools for rare cell profiling, immune landscape dissection, and studies of dynamic cell state transitions in complex biological systems.

**Challenges and Perspectives**

In this perspective, we reviewed the application of ferrofluids-based microfluidics on label-free, precise manipulation and analysis of bioparticles across a broad size spectrum (from micron-scale cells to submicron vesicles) and wide physical traits (size, shape, and antigen-binding capacity). These microfluidic platforms have enabled high-resolution bioparticle separation and cell phenotyping based on physical and mechanical properties, providing innovative approaches for rare cell isolation, single-cell analysis, and high-resolution phenotype profiling. However, technical and translational challenges must be addressed before these methods can be broadly implemented in biological and clinical applications.

(1) **Biocompatibility of Ferrofluids.** The biocompatibility of ferrofluids continues to be a major concern when applied to live cells or functional downstream assays. Surfactants and stabilizers in ferrofluids and cellular updates of nanoparticles can potentially interfere with cell viability, signaling, and gene expression. The limited biocompatibility limits the assay time using ferrofluids, as the ferrofluids need to be removed as soon as possible. While our lab has developed biocompatible ferrofluids that have minimized impact on cell viability, proliferation, and motility-related function, further improvements are needed to support long-term cell screening, live-cell recovery, and various functional assays post-separation. To enhance biocompatibility, ferrofluids can be synthesized and optimized using surface-coated magnetic nanoparticles (e.g., lipid bilayer coating) and physiologically compatible carriers (e.g., cell culture medium). Reducing surfactant toxicity, optimizing magnetic nanoparticle size, and enhancing coating stability are key to minimizing cellular uptake and preserving cell functions during ferrofluids exposure.

(2) **Scalability and Throughput.** High throughput and efficient processing of large volumes of biological samples (e.g., whole blood, lavage fluid, food extracts) require ferrohydrodynamic systems that balance magnetic buoyancy force precision with robust processing capacity. While platforms such as inertial-FCS have demonstrated promising throughput levels (>60 mL/h human blood), further optimization is needed to handle high-cellularity samples like leukapheresis products, which can contain ~$10^9$ WBCs. Additionally, current ferrohydrodynamic systems exhibit limited throughput for submicron particle enrichment (e.g., exosomes), restricting their application for milliliter- to liter-scale sample volumes. To overcome these limitations, further optimization is needed to enable continuous processing, minimize sample loss, and maintain cell manipulation performance over long processing times. The throughput can be increased by enhancing magnetic flux density & density gradient using multiple magnet arrays and magnet pole configurations. Geometric optimization of microfluidic channels, or the use of parallel channels, can also help balance magnetic exposure time with increased flow rates. However, these strategies introduce new design challenges, including magnet-channel alignment, fluidic uniformity, and device complexity. Moreover, current separation performance is highly dependent on the efficiency of sample focusing. Integrating ferrohydrodynamic systems with effective flow-focusing strategies, such as inertial focusing, offers a scalable solution that supports high flow rates while preserving separation resolution. Continued efforts to co-optimize magnetic force fields, channel design, magnet & channel alignment via 3D printing techniques, and focusing dynamics will be essential for advancing ferrofluids-based separation technologies toward clinical-scale processing.

(3) **Ferrofluids Nanoparticles Depletion.** Post-depletion of residual ferrofluids nanoparticles is challenging and essential, especially for sensitive downstream analysis, including sequencing, mass spectrometry, or therapeutic use. While conventional centrifugation is suitable for micron-scale particles, it is ineffective for submicron targets, where size and density overlap with ferrofluids components. Ultracentrifugation can deplete nanoparticle residues, but it adds processing time and cost and often results in sample loss. To address this, we previously applied a strategy that disrupts ferrofluids stability by modifying the carrier solution's pH. By introducing sodium bicarbonate to raise the medium pH to ~7.5–8.0, we induced aggregation and sedimentation of ferrofluids nanoparticles, enabling their removal via low-speed centrifugation (1000g, 5 min)[26]. This approach was successfully applied to exosome purification; however, its impact on exosomal protein and miRNA integrity remains to be systematically evaluated. New methods are still needed to improve nanoparticle depletion from submicron samples while preserving biological content. One potential strategy involves the use of magnetic field-assisted depletion, in which residual ferrofluids nanoparticles are selectively pulled to one side of the container or microchannel using an external gradient field, followed by decanting or directional flow to separate submicron targets. Another approach could involve stimuli-responsive surfactants or coatings, such as temperature-sensitive polymers (e.g., PNIPAM[77]), that can be triggered to induce reversible aggregation of ferrofluids nanoparticles, facilitating their separation under mild conditions without compromising target particle quality.

(4) **Extended Ferrofluids-Based Techniques.** Beyond CTCs and exosome separation, ferrofluids-based platforms hold exciting potential for isolating a broader range of bioparticles, including dendritic cells, microglia, and subcellular organelles such as mitochondria, endosomes, and lysosomes. New separation strategies are of great clinical and biological interest but remain unexplored in the ferrohydrodynamic separation field. Additionally, there are promising opportunities to extend ferrofluids-based techniques beyond traditional size-based separation into less-explored phenotypic dimensions. These include cell deformability, nuclear-to-cytoplasmic ratio, cellular or subcellular granularity, and dynamic biophysical responses under external stimulation. We have noted the increasing influence of machine learning and deep learning in guiding microfluidic design and phenotyping strategies. Looking forward, we anticipate significant potential in integrating artificial intelligence with ferrofluids-enabled separation and profiling platforms. The combination of real-time imaging, label-free detection techniques (e.g., SERS imaging), and AI-driven data analysis could enable multiparametric, high-content phenotyping at the single-cell level, opening new avenues for decoding complex cell states and dynamic transitions in heterogeneous biological systems.

In summary, ferrofluids-based microfluidics offers a powerful and versatile platform for physical trait-based bioparticle separation and cell phenotyping. Addressing current challenges in biocompatibility, scalability, and nanoparticle removal will be crucial for unlocking its full potential. With continued innovation, these technologies have the potential to revolutionize precision diagnostics, rare cell analysis, immune profiling, and translational research across biomedicine and biotechnology.

**Acknowledgements:** This work was funded by the National Institutes of Health (NIH) 1R01GM16324.

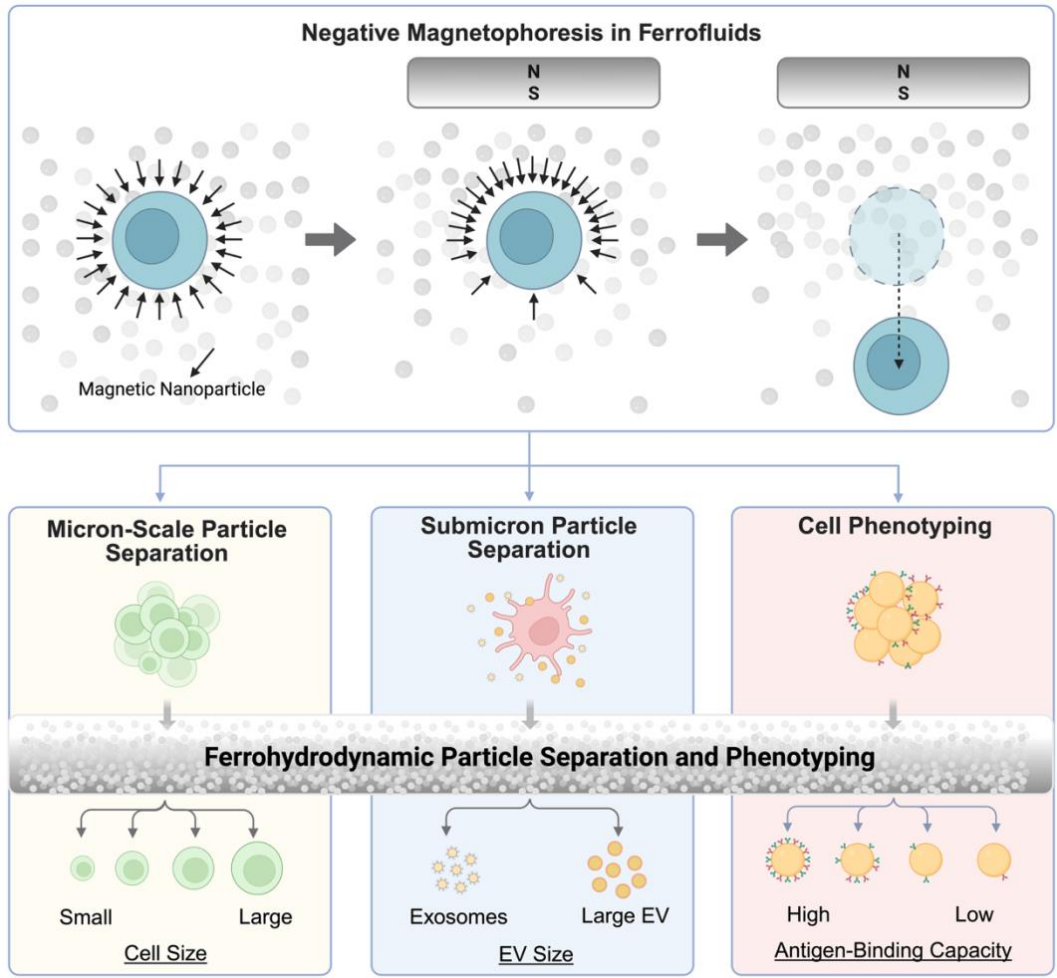

*Figure 1: Overview and applications of negative magnetophoresis on micron-scale particle separation, submicron particle separation, and cell phenotyping using ferrofluids.* Negative magnetophoresis in ferrofluids enables label-free particle separation and phenotyping across scales. Diamagnetic particles suspended in ferrofluids experience a repulsive force under an external magnetic field, resulting in controlled migration away from high-field regions. This principle underlies a versatile platform for ferrohydrodynamic separation. Top: schematic of the physical mechanism, where the magnetic susceptibility contrast between the particle and the ferrofluid drives displacement under a magnetic gradient. Bottom: applications span from micron-scale cell sorting by size, submicron vesicle separation based on extracellular vesicle (EV) size, to phenotypic sorting of cells labeled with magnetic nanoparticles targeting surface antigens. Created with BioRender.com released under a Creative Commons Attribution-NonCommercial-NoDerivs 4.0 International license.